\documentclass[conference]{IEEEtran}

\ifCLASSINFOpdf

\else
\fi

\usepackage{url}
\usepackage[british]{babel}
\usepackage{hhline}
\usepackage{multirow}
\usepackage{graphicx}
\usepackage{epsfig}
\usepackage{ifmtarg}
\usepackage[font=small,labelsep=none]{caption}
\usepackage[font=footnotesize]{caption}
\usepackage{footnote}
\usepackage{cite}
\usepackage{amssymb}
\usepackage{amsthm}
\usepackage[fleqn]{amsmath}
\usepackage{subfig}
\usepackage{mathtools}
\usepackage{amsfonts}
\usepackage{epstopdf}
\usepackage{epsfig}
\usepackage[linesnumbered,ruled]{algorithm2e}
\usepackage{algorithmic}
\usepackage[font=small,labelsep=none]{caption}
\usepackage[font=footnotesize]{caption}
\usepackage[left=0.631in, right=0.631in, top=0.7in, bottom=1.025in]{geometry}
\begin{document}
\title{Short-term and Long-term Cell Outage Compensation Using UAVs in 5G Networks}
\author{
\IEEEauthorblockN{\large Mohamed Y. Selim$^{\text{1}}$, Ahmad Alsharoa$^{\text{2}}$ and Ahmed E. Kamal$^{\text{1}}$} \\
\vspace{-.15 in}
\small
\IEEEauthorblockA{$^{\text{1}}$Iowa State University, Iowa, USA, Email: \{myoussef, kamal\}@iastate.edu}\\ \vspace{-.15 in}
\IEEEauthorblockA{$^{\text{2}}$University of Central Florida, Florida, USA, Email: ahmad.al-sharoa@ucf.edu}\\ \vspace{-.15 in}
%\IEEEauthorblockA{$^{\text{5}}$King Saud University, Riyadh, Saudi Arabia, Email: malnuem@ksu.edu.sa}\\ \vspace{-.5 in}
%\normalsize
\vspace{-1cm}
}

%%%%%%%%%%%%%%%%%%%%%%%%%%%%%%%%%%%%%%%%%%%%%%%%%%%%%%%%%%%%%%%%%%%%%%%%%%%%%%%%%%%%%%%%%%%%%%%%%%%%%%%%%%%%%%%%%%%%%%%%%%%%%%%%%%%%%%%%%%%%%%%%%%%%%%%
%%%%%%%%%%%%%%%%%%%%%%%%%%%%%%%%%%%%%%%%%%%%%%%%%%%%%%%%%%%%%%%%%%%%%%%%%%%%%%%%%%%%%%%%%%%%%%%%%%%%%%%%%%%%%%%%%%%%%%%%%%%%%%%%%%%%%%%%%%%%%%%%%%%%%%%
%ADD THE FOLLOWING TO THE PAPER BEFORE THE CAMERA READY
%centralized at the tethered BD
%Reference UK paper

%%%%%%%%%%%%%%%%%%%%%%%%%%%%%%%%%%%%%%%%%%%%%%%%%%%%%%%%%%%%%%%%%%%%%%%%%%%%%%%%%%%%%%%%%%%%%%%%%%%%%%%%%%%%%%%%%%%%%%%%%%%%%%%%%%%%%%%%%%%%%%%%%%%%%%%
%%%%%%%%%%%%%%%%%%%%%%%%%%%%%%%%%%%%%%%%%%%%%%%%%%%%%%%%%%%%%%%%%%%%%%%%%%%%%%%%%%%%%%%%%%%%%%%%%%%%%%%%%%%%%%%%%%%%%%%%%%%%%%%%%%%%%%%%%%%%%%%%%%%%%%%
\maketitle
%\pagestyle{plain}
%\thispagestyle{plain}
%\pagenumbering{arabic}
%\thispagestyle{empty}
%\pagestyle{empty}

\begin{abstract}
%\boldmath
The use of Unmanned Aerial Vehicles (UAVs) has gained interest in wireless networks for its many uses and advantages such as rapid deployment and multi-purpose functionality. This is why wide deployment of UAVs has the potential to be integrated in the upcoming 5G standard. They can be used as flying base-stations, which can be deployed in case of ground Base-Stations (GBSs) failures. Such failures can be short-term or long-term. Based on the type and duration of the failure, we propose a framework that uses drones or helikites to mitigate GBS failures. Our proposed short-term and long-term cell outage compensation framework aims to mitigate the effect of the failure of any GBS in 5G networks. Within our framework, outage compensation is done with the assistance of sky BSs (UAVs). An optimization problem is formulated to jointly minimize communication power of the UAVs and maximize the minimum rates of the Users' Equipment (UEs) affected by the failure. Also, the optimal placement of the UAVs is determined. Simulation results show that the proposed framework guarantees the minimum quality of service for each UE in addition to minimizing the UAVs' consumed energy.
\end{abstract}
\IEEEpeerreviewmaketitle
\section{Introduction}

Unmanned Aerial Vehicles (UAVs) enabled communications is considered as a strong candidate to be used in 5G networks. Indeed, UAVs enabled communications offers an encouraging solution to provide wireless connectivity for devices without coverage due to, e.g., severe shadowing by urban or mountainous terrains, unexpected failures, or damage to the communication infrastructure due to malicious or natural causes \cite{Vatta}.

% This paragraph from comm magazine May 2016 paper
Drones are a special type of UAVs that are popular for remote sensing and surveillance. Recently drones were used by Nokia to provide connectivity to smart cities (www.nokia.com). Although drones are very popular in UAV-based communications, there are other types of UAVs which are strong candidates to be used as flying Base-Stations (BSs) in 5G. The most relevant and well-known UAV types are:

$\bullet$\textbf{Drones}:
are a special type of UAVs that are used in many applications nowadays and they are gaining increasing popularity in information technology applications due to their high flexibility for on-demand deployments. Due to their relatively low capacity, both in terms of payload and autonomy, they are generally restricted to low altitudes.

$\bullet$ \textbf{Aircrafts}:
powered by fuel or batteries, are capable of remaining aloft for several days. This category of UAVs possesses favorable features such as low-power and energy-efficient lightweight structures with sufficient payload capacity which allow efficient trajectory management and positioning tools \cite{Xiongfeng}. This type of UAV is not used widely in the cellular networks since it can't be stored near to Ground BSs (GBSs).

$\bullet$ \textbf{Airships}:
These UAVs which utilize lighter gas to float in air are classified as aerostatic platforms. Airships are much more flexible in terms of weight, size and power consumption. They have been designed to fly up to 20 km. They are capable of staying in the air for long periods of time, which may be even months. A well-know example of already deployed airships is project Loon powered by Google (www.google.com/loon).

$\bullet$ \textbf{Helikites}:
The Helikite exploits both wind and helium for its lift. The aerodynamic lift is essential to combat the wind meanwhile its power consumption is very low. Helikites are very popular low altitude platforms operable independent of weather conditions and can stay in air for a few weeks \cite{Bucaille}.

Table 1 summarizes the capabilities of the aforementioned UAVs \cite{Table}.

\small
\begin{table}[h!]
\caption{: Comparing different types of UAVs}\label{epsilontable}
\centering
\addtolength{\tabcolsep}{-4pt}\begin{tabular}{||c||c||c||c||c||c||c||c||c||c||c||} % centered columns (4 columns)
%\toprule %inserts double horizontal lines
\hline
 \textbf{UAVs Capabilities} & \textbf{Drones} & \textbf{Aircraft} & \textbf{Airship} & \textbf{Helikite} \\ [0.5ex] % inserts table
 \hline \hline
 \vspace{.01 in}
%heading
%\midrule % inserts single horizontal line
\textbf{High Payload} & Based on size & Yes & Yes & Yes\\
\hline
\vspace{.01 in}
\textbf{Moving Coverage} & Yes & Yes & Yes & No\\
\hline
\vspace{.01 in}
\textbf{Instant Deployment} & Yes & Yes & No & Yes\\
\hline
\vspace{.01 in}
\textbf{Weather Resistance} & No & Yes & No & Yes\\
\hline
\vspace{.01 in}
\textbf{Easily Handled} & Yes & No & No & Based on size\\
\hline
\vspace{.01 in}
\textbf{Power Consumption} & High & High & High & Low\\
\hline
%\bottomrule %inserts single line
\end{tabular}
\label{table1}
\end{table}
\normalsize
%%%%%%%%%%%%%%%%%%%%%%%%%%%%%%%%%%%%%%%%%%%%%%%%%%%%%%%%%%%%%%%%%%%%%%%%%%%%%%%%%%%%%%%%%%%%%%%%%%%%%%%%%%%%%%%%%%%%%%%%%%%%%%%%%%%%%%%%%%%%%%%%%%%%%%%%%%%%%%%%%%%%%%%%%%%%%%%%%%%%%%%

Cellular systems are prone to failures, and the most critical domain for fault management is the radio access network. Operator's revenue losses occurs when at least one BS fails for a short period of time. Longer failures bias users to switch to competitors which results in permanent revenue losses.

Self-organizing Networks (SONs) are used to leapfrog the overall performance of the network to a higher level of automated operation in the 5G network management. This concept has been introduced by 3GPP in Release 8 and it has been expanding across subsequent releases \cite{3GPP}. SON defines three areas: self-configuration (plug and play network elements), self-optimization (optimize network elements and parameters) and self-healing (automatically detect and mitigate failures) \cite{Imran}.

Self-healing is done in two steps: Cell Outage Detection (COD) and Cell Outage Compensation (COC). The COD is to detect and classify failures, while minimizing the detection time. The COC aims to mitigate the effect of the failure. If the failure time exceeds a certain threshold, it is considered as a long-term failure otherwise it is considered short-term \cite{bookchapter}.

When a failure occurs to any GBS, the conventional COC technique is to adjust the neighboring BSs' antenna tilt and power to serve the users of the failed BS. This technique is very fast and guarantees minimum Quality of Service (QoS) to the users given a failed BS. However, the disadvantage of this technique is that the users of the neighboring BSs will be affected by the change in their BS's antenna configuration.

We proposed a solution for this problem that mainly depends on using UAVs as flying BSs. These UAVs are initially co-located with GBSs and ready to fly when needed. When the failure occurs, UAVs will fly to their initial positions to start compensating UEs of the failed BS. During this flying time, the conventional self-healing technique is used to serve those UEs until UAVs reach their predetermined locations. When the UAVs reach these pre-computed locations, the neighboring BSs  return to serve their own users only.

Based on the comparison presented in Table 1, we propose to use Drone BSs (DBSs) in healing short-term failures since they have the important feature of instant deployment, especially if the network operator already placed ready-to-fly drones at each cell site. For long-term failures, the helikite is proposed to heal the failed BS since it flies at low altitudes and its flying power consumption is the lowest compared to other types of UAVs for long flying periods. Weather conditions must be considered when using DBSs. Hence, if weather conditions are not suitable for DBSs to aviate, it is recommended to use helikites even if we are dealing with a short-term failure.

%The mobile BSs will use the same resources of the failed BS. This is because the failed BS will not be able to serve its users until it is repaired.

%add more from paper 1 and also from walid saad paper

Although there has been a recognized amount of work on using DBSs in cellular networks, using DBSs in self-healing is still at its infancy. The authors in \cite{Lyu} presented a novel idea of offloading the traffic of UEs suffering from degraded service at the GBS cell edge. They jointly optimizing the UAV's trajectory, as well as the user partitioning between the UAV and GBS. In \cite{Rohde}, the positioning of aerial relays is discussed to compensate cell outage and cell overload. The authors in \cite{Merwaday} show the improvement in coverage by assisting the network with DBSs at a certain altitude during BS failure.

The authors in \cite{Selimm} present a novel COC framework to mitigate the effect of the failure of any BS in 5G networks using both UAVs and GBSs. They showed that their proposed hybrid approach outperforms the conventional COC approach.
In \cite{Alzenad}, a vertical backhaul/fronthaul
framework is suggested for transporting the traffic between
the access and core networks in a typical HetNet through
free space optical links.

%%%%%%%%%%%%%%%%%%%%%%%%%%%%%%%%%%%%%%%%%%%%%%%%%%%%%%%%%%%%%%%%%%%%%%%%%%%%%%%%%%%%%%%%%%%%%%%%%%%%%
%%%%%%%%%%%Add A reference for Rui Zhang and the moderator of the workshop%%%%%%%%%%%%%%%%%%%%%%%%%%%
%%%%%%%%%%%%%%%%%%%%%%%%%%%%%%%%%%%%%%%%%%%%%%%%%%%%%%%%%%%%%%%%%%%%%%%%%%%%%%%%%%%%%%%%%%%%%%%%%%%%%

%%%%%%%%%%%%%%%%%%%%%%%%%%%%%%%%%%%%%%%%%%%%%%%%%%%%%%%%%%%%%%%%%%%%%%%%%%%%%%%%%%%%%%%%%%%%%%%%%%%%%
%%%%%%%%%%%This paper is organized as follows %%%%%%%%%%%%%%%%%%%%%%%%%%%%%%%%%%%%%%%%%%%%%%%%%%%%%%%
%%%%%%%%%%%%%%%%%%%%%%%%%%%%%%%%%%%%%%%%%%%%%%%%%%%%%%%%%%%%%%%%%%%%%%%%%%%%%%%%%%%%%%%%%%%%%%%%%%%%%

%\subsection{Our Contributions}
%The main contributions of this paper are summarized below:

%\begin{itemize}

%\item Mitigating the failure effect in 5G networks using drone-based communication by introducing a full and novel self-healing framework.

%\item Optimizing the sum rate of UEs under the failed BS by finding the optimal 3D positioning of the healing DBSs.

%\item Using conventional self-healing technique to serve UEs under the failed BS until the DBSs arrive to their initial positions and begin serving these UEs.

%\item Reducing DBS's energy consumption which results in extending it's flying and serving time by using the proposed concept of Partial Cloud Radio Access Network (PCRAN).

%\item Proposing a solution for the unexplored issue of replacing DBSs during the healing process without interrupting the connectivity of the active UEs.

%\end{itemize}

\section{System Architecture}

\begin{figure}
            \centering
        \includegraphics[width=3in, height=1.8in]{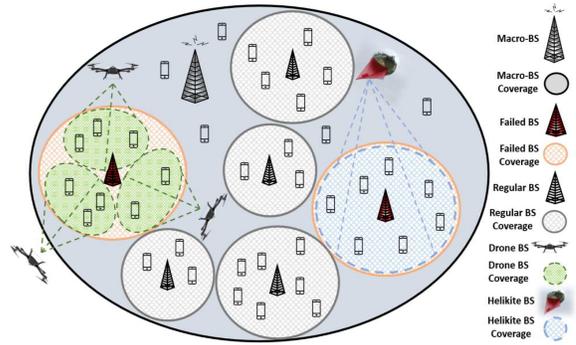}
        \caption{\,: System model during failure.}
        \label{Sysmodel}
\end{figure}

We consider a downlink heterogeneous network consisting of a Macro-Base Station overlaying number of Small BSs (SBSs).  Fig. \ref{Sysmodel} shows the network architecture during the failure of two SBSs. In this figure we show a short-term failure which is mitigated using three DBSs and a long-term failure which is mitigated using one helikite.

The set $\mathcal{U}=\{1,2,\dots,U\}$ denotes the set of active UEs under the failed BS and they are at known locations where the horizontal coordinates of each UE $u$ are fixed at $\mathbf{g}_u=[x_u,y_u]^T \in \mathbb{R}^{2\text{x}1}$, $u \in U$, assuming that all UEs are at zero altitude. The set $\mathcal{D}=\{1,2,\dots,D\}$ denotes the set of DBSs used to heal the failed BS where all DBSs are assumed to navigate at a fixed altitude $h_d$ and the horizontal coordinates of DBS $d$ are denoted by $\mathbf{J}_d=[x_d,y_d]^T \in \mathbb{R}^{2\text{x}1}$.

%%%%%%%%%%%%%%%%%%%%%%%%%%%%%%%%%%%%%%%%%%%%%%%%%%%%%%%%%%%%%%%%%%%%%%%%%%%%%%%%%%%%%%%%%%%%%%%%%%%%%%%%%%%%%%%%%%%%%%%%%%%%%%%%%%%%%%%%%%%%%%

We denote that DBS $d$ is communicating with UE $u$ using resource block $m$ by the binary variable $\Phi_{u,d}^m$ which acts as a decision variable in our problem formulation. We denote by $\psi_{u,d}$ the binary association between DBS $d$ and UE $u$.

Assume that the DBS-UE communication channels are dominated by LoS links. Though simplified, the LoS model offers a good approximation for practical Drone-UE channels and enables us to investigate the main objective of the optimization problem presented later. Under the LoS model, the Drone-UE channel power gain follows the free space path loss model which is determined mainly by the DBS-UE distance. Given that $\mathbf{J}_d$ and $\mathbf{g}_u$ are the coordinates of DBS $d$ and UE $u$ in the horizontal plane, respectively, then the distance from DBS $d$ to UE $u$ can be expressed as: $\delta_{u,d}=\sqrt{h_d^2+||\mathbf{J}_d-\mathbf{g}_u||^2}$ .

\subsection{DBS Channel and Achievable Rate Models}

For simplicity, we assume that the communication links DBS-UE are dominated by the LoS links where the channel quality depends only on the distance between the DBS and the UE. Under this LoS model, the DBS-UE channel power gain mainly follows the free space path loss model which is given as follows:

\vspace*{-0.1in}
\small
\begin{equation}\label{pathlssd}
  %\Gamma_{u,d}=\rho_{o} (\delta_0/\delta_{u,d})^{2}=\frac{\rho_{o}}{h^2+||\mathbf{J}_d-\mathbf{g}_u||^2}
  \Gamma_{u,d}=\rho_{o} (\delta_0/\delta_{u,d})^{2}=\rho_{o}/\big({h^2+||\mathbf{J}_d-\mathbf{g}_u||^2}\big)
\end{equation}
\normalsize

where $\rho_{o}$ is a unitless constant that depends on the antenna characteristics and frequency, and is measured at the reference distance $\delta_0=\text{1~m}$.

Let $\mathcal{M}=\{1,2,\dots,M\}$ be the set of sub-channels that each DBS can use during the self-healing process. These sub-channels will be further divided and allocated to the UEs associated with each DBS. Each DBS, $d$, transmits to each UE, $u$, with a per sub-channel transmit power $p_{u,d}^m$. If sub-channel $m$ is not assigned to DBS $d$ then $p_{u,d}^m$ will equal to zero. For simplicity, we assume that there is no interference between the DBS tier and the GBS tier which means that each of them is using different sets of sub-channel. However, we consider the interference between different DBSs. Hence, the received Signal to interference plus Noise Ratio between DBS $d$ and UE $u$ per sub-channel $m$ can be expressed as:

\small
\begin{equation}\label{SINRd}
  \gamma_{u,d}^m= \frac{p_{u,d}^m~\Gamma_{u,d}^m}{\sum\limits_{\substack{i\in\mathcal{U} \\ i\neq u}}\sum\limits_{\substack{j\in\mathcal{D}}} p_{i,j}^m \Gamma_{u,j}^m + \sigma^2} = \frac{\frac{p_{u,d}^m ~\rho_{o}}{h^2+||\mathbf{J}_d-\mathbf{g}_u||^2}}{\sum\limits_{\substack{i\in\mathcal{U} \\ i\neq u}}\sum\limits_{\substack{j\in\mathcal{D}}} \frac{p_{i,j}^m ~\rho_{o}}{h^2+||\mathbf{J}_j-\mathbf{g}_u||^2}  + \sigma^2}
\end{equation}
\normalsize

%define epsilon and zeta
where $\sigma^2$ is the power of the Additive White Gaussian Noise at the receiver. The first term in the denominator of equation (\ref{SINRd}) represents the co-channel interference caused by the transmissions of all other DBSs on the same sub-channel $m$, respectively. %Thus the achievable per sub-channel rate of UE $u$  connected to DBS $d$ is $R_{u,d}^m= \text{log}_\text{2}(1+\gamma_{u,d}^m)$  bps/Hz.

% I need to define
%%%%%%%%%%%%%%%%%%%%%%%%%%%%%%%%%%%%%%%%%%%%%%%%%%%%%%%%%%%%%%%%%%%%%%%%%%%%%%%%%%%%%%%%%%%%%%%%%%%%%%%%%%%%%%%%%%%%%%%%%%%%%%%%%%%%%%%%%%%%%

Accordingly, the achievable per sub-channel downlink rate from DBS $d$ to UE $u$ is given by:

\begin{equation}
  R_{u,d}^m=\text{log}_{\text{2}}(1+\gamma_{u,d}^m)
\end{equation}

%from Ahmad rephrase
\subsection{UAV Power Model}
%talk about the different power models to different UAVs and this is why we didn't include the power model in the optimization problem and we only included the communication power.

Since the proposed framework allows different types of UAVs to compensate the failure based on the type of failure (short-term or long-term),two types of UAVs are proposed to be used in this self-healing framework; Drones and Helikite. The operation power of Drones is very high due to the hovering and hardware power \cite{wcnc}. However, the operation power of Helikites is much lower since its weight is lifted by the helium and additional power is consumed only to sustain the location of the Helikite. From minimizing the consumed power point of view, we assign Drones to short-term healing and Helikites to long-term healing. This is why in the formulated optimization problem we consider minimizing the downlink power $p_{u,d}^m$ of the UAV regardless its type, i.e., Drone or Helikite.

\section{Problem Formulation}

In this section, we formulate an optimization problem aiming to maximize the minimum achievable rate of the UEs under the failed GBS and meanwhile minimizing the transmission power of the UAV used, i.e., either the DBS or the helikite. The number of UAVs used to heal a failed GBS is based on the coverage area of the failed GBS and the type of the UAVs used. The optimization problem formulation is given by:

%(1/P^{\text{max}}*|\mathcal{D}|)
%(1/R^{\text{th})

\small
\begin{subequations}
\begin{align}
&\hspace{-0.5cm} (\textbf{P1}):\underset{{\mathbf{J}}, {\Phi}, {\Psi} , {\textbf{p}}}{\text{maximize}}  \frac{\Omega}{R^{\text{th}}} - \frac{1}{P^{\text{max}}*|\mathcal{D}|} \sum_{\textit{d}}\sum_{\textit{u}}\sum_{\textit{m}} \psi_{u,d} \Phi_{u,d}^m p_{u,d}^m    \label{of}\\
&\hspace{-0.5cm} \text{subject to:}\nonumber\\
&\hspace{-0.5cm} \Omega \leq \sum_{d} \sum_{m} \psi_{u,d} \Phi_{u,d}^m R_{u,d}^m, \quad \forall~u      \label{maxmin}\\
&\hspace{-0.5cm}  \sum_{d} \sum_{m} \psi_{u,d}\Phi_{u,d}^m R_{u,d}^m \geq R^{\text{th}}, \quad \forall~u \label{minR}\\
&\hspace{-0.5cm} \sum_{d}\sum_{m}\Phi_{u,d}^m \geq 1, \quad \forall~ u \label{phi}\\
&\hspace{-0.5cm}\mathbf{J}_d^{\text{min}} \leq \mathbf{J}_d \leq \mathbf{J}_d^{\text{max}}, \quad \forall~ d   \label{Coordinates}\\
&\hspace{-0.5cm} \sum_{u} \sum_{m} p_{u,d}^m \leq P^{\text{max}} \label{maxpower}, \quad \forall~ d\\
&\hspace{-0.5cm} p_{u,d}^m \geq 0 , \quad \forall~ u,d,m  \label{nonnegativep}\\
&\hspace{-0.5cm} \sum_{d} \psi_{u,d}=1 \quad \forall~ u     \label{AssociationC}\\
&\hspace{-0.5cm} \Phi_{u,d}^m \in \{0,1\}  \quad \forall~ d,  \label{energyC}
\end{align}
\end{subequations}
\normalsize

Eq. (\ref{of}) represents the objective function where the first term is  maximizing the minimum achievable rate of the UEs originally served by the failed GBS where $\Omega$ is an auxiliary continuous variable used to represent the maximization of the minimum achievable rate of the UEs. The second term aims to minimize the sum of the downlink transmission power of all UAVs given that $\Phi_{u,d}^m$ is the resource allocation binary variable which will equal to zero if sub-channel $m$ is not used for the downlink transmission between DBS $d$ and UE $u$. Constraint (\ref{maxmin}) is the mathematical representation of max-min where we are trying to maximize $\Omega$ which is less than or equal the achievable rate of all UEs, i.e., maximizing the minimum rate. Constraint (\ref{minR}) represents the QoS constraint on the rate of each UE, $u$, where $R^{th}$ is the threshold rate. In constraint (\ref{phi}), each UE is forced to acquire at least one sub-channel. Constraint (\ref{Coordinates}) is used to limit the 2D coordinates of DBS $d$ where $\mathbf{J}_d^{\text{min}}=[x_d^{\text{min}},y_d^{\text{min}}]^T$ and $\mathbf{J}_d^{\text{max}}=[x_d^{\text{max}},y_d^{\text{max}}]^T$. The maximum and minimum power limits are presented in constraints (\ref{maxpower}) and (\ref{nonnegativep}). Constraint (\ref{AssociationC}) enforce each user to associated with only one DBS.

$\bf{P1}$ is not easy to solve due to the following: 1) the decision variables $\Phi_{u,d}^m$ and $\psi_{u,d}$ are binary and thus the objective function (\ref{of}) and constraints (\ref{maxmin})-(\ref{phi}) involve binary constraints which makes solving it a hard problem. 2) Even if we fixed the decision variables, constraints (\ref{maxmin}) and (\ref{minR}) are still non-convex with respect to DBS coordinates variable $\mathbf{J}_d$ and downlink power, $p_{u,d}^m$. Therefore, problem (\ref{of}) is mixed-integer non-linear non-convex problem, which is difficult to be solved optimally.

To make $\bf{P1}$ more tracktable, we reformulate $\bf{P1}$ as follows:

\small
\begin{subequations}
\begin{align}
&\hspace{-0.5cm} (\textbf{P2}): \underset{{\mathbf{J}}, {\Phi}, {\Psi}, {\textbf{p}}}{\text{maximize}}  \quad \frac{\Omega}{R^{\text{th}}} - \frac{1}{P^{\text{max}}*|\mathcal{D}|} \sum_{\textit{d}}\sum_{\textit{u}}\sum_{\textit{m}} p_{u,d}^m    \label{of1}\\
&\hspace{-0.5cm} \text{subject to:}\nonumber\\
&\hspace{-0.5cm} ~~~~~~~~~~~~~~~~~~~~~~~~~\text{Constraints (\ref{phi})~-~(\ref{energyC})} \nonumber\\
&\hspace{-0.5cm} \Omega \geq R^{\text{th}} \label{minrate1}\\
&\hspace{-0.5cm} \sum_{d}\sum_{m} \text{log}_{\text{2}}(1+\frac{p_{u,d}^m~\Gamma_{u,d}^m}{\sum\limits_{\substack{i\in\mathcal{U} \\ i\neq u}}\sum\limits_{\substack{j\in\mathcal{D}}} p_{i,j}^m \Gamma_{u,j}^m + \sigma^2}) \geq \Omega, \quad \forall~ u  \label{maxmin1}\\
&\hspace{-0.5cm} p_{u,d}^m \leq \psi_{u,d} \Phi_{u,d}^m P^{\text{max}}, \quad \forall~ u,d,m \label{powerphi1}
\end{align}
\end{subequations}
\normalsize

The main difference between $\bf{P2}$ and $\bf{P1}$ is that we added constraint (\ref{powerphi1}) to $\bf{P2}$ in addition to rewriting constraints (\ref{maxmin}) and (\ref{minR}). Constraint (\ref{powerphi1}) is used mainly to force $p_{u,d}^m$ to equal to zero if $\Phi_{u,d}^m$ and/or $\psi_{u,d}$ equal to zero. Consequently, there is no need to multiply the term $\psi_{u,d}\Phi_{u,d}^m$ by $p_{u,d}^m$ as done in the objective function of $\bf{P1}$. The same concept apply to constraints (\ref{maxmin}) and (\ref{minR}) where when $\psi_{u,d}$ or $\Phi_{u,d}^m$ equals to zero then $p_{u,d}^m$ will equal to zero which consequently will result in $R_{u,d}^m$ equals to zero. Similarly, constraint (\ref{powerphi1}) is used to eliminate $\psi_{u,d}$ from constraints (\ref{maxmin}) and (\ref{minR}). Since $\Omega$ main purpose is to maximize the minimum achievable rate, then using the constraint $\Omega \geq R^{\text{th}}$ is doing the same purpose of constraint (\ref{minR}). However in this case, we are guaranteeing that the minimum rate is greater than or equal a certain threshold.

Constraint (\ref{powerphi1}) is non-linear due to the multiplication of the two decision variables $\psi_{u,d}$ and $\Phi_{u,d}^m$. This constraint can be exactly linearized, i.e., without any approximation, by replacing it by the following three constraints:

\small
\begin{subequations}
\begin{align}
&\hspace{-0.5cm} p_{u,d}^m \leq \psi_{u,d} P^{\text{max}},  \quad\quad\quad\quad\quad\quad\quad\quad\quad \forall~ u,d,m \label{linear1}\\
&\hspace{-0.5cm} p_{u,d}^m \leq \Phi_{u,d}^m P^{\text{max}}, \quad\quad\quad\quad\quad\quad\quad\quad\quad \forall~ u,d,m \label{linear2}\\
&\hspace{-0.5cm} p_{u,d}^m \geq (\psi_{u,d}+\Phi_{u,d}^m-1)P^{\text{max}}, ~\quad\quad\quad \forall~ u,d,m \label{linear3}
\end{align}
\end{subequations}
\normalsize

$\bf{P2}$ is still not easy to solve due to the binary variables $\Phi_{u,d}^m$ and $psi_{u,d}$ and the non-linearity in constraint (\ref{maxmin1}). In addition, constraint (\ref{maxmin1}) has inside the logarithmic term two variables one in the numerator and the other in the denominator. However, $\bf{P2}$ is more tracktable and easier to solve than $\bf{P1}$ given that $\bf{P2}$ is a new version of $\bf{P1}$ without any approximation.

\section{The Proposed Solution}

In general, $\bf{P2}$ has no standard method for solving it efficiently. In the following, we propose an efficient iterative algorithm for solving $\bf{P2}$. Specifically, for a given coordinate $\mathbf{J}_d$, we optimize the decision variables $\Phi_{u,d}^m$ and $\psi_{u,d}$ and the continuous variable $p_{u,d}^m$ based on the Successive Convex Approximation (SCA) technique \cite{SCAICC}. Then for a given resource allocation and power, we find the near optimal coordinates using heuristic iterative technique. Finally, a joint iterative algorithm is proposed to solve $\bf{P2}$ efficiently.

\subsection{UAV Downlink Power and Resource Allocation}\label{uav1}

For any given coordinates, $\mathbf{J}_d$, the UAV downlink power and resource allocation of $\bf{P2}$ can be optimized by solving the following problem:

\small
\begin{align}
&\hspace{-0.5cm} (\textbf{P3}): \underset{{\Phi_{\textit{u},\textit{d},\textit{m}}}, {\textit{p}_{\textit{u},\textit{d},\textit{m}}}}{\text{maximize}}  \quad \frac{\Omega}{R^{\text{th}}} - \frac{1}{P^{\text{max}}*|\mathcal{D}|} \sum_{\textit{d}}\sum_{\textit{u}}\sum_{\textit{m}} \textit{p}_{\textit{u},\textit{d},\textit{m}}    \label{of2}\\
&\hspace{-0.5cm} \text{subject to:}\nonumber\\
&\hspace{-0.5cm} ~~~~~~~~~~\text{Constraints (\ref{phi}), (\ref{maxpower})~-~(\ref{energyC}), (\ref{minrate1})~-~(\ref{maxmin1}), (\ref{linear1})~-~(\ref{linear3})} \nonumber
\end{align}
\normalsize

$\bf{P3}$ is a non-convex optimization problem due to the non-convex constraint (\ref{maxmin1}). Based on the mathematical manipulation presented in \cite{RuiZhangTraj}, this constraint can be rewritten as:

\small
\begin{align}
&\sum_{m} \Big(\underbrace{\text{log}_{\text{2}}\big(\sum\limits_{\substack{i\in\mathcal{U}}}\sum\limits_{\substack{j\in\mathcal{D}}} p_{i,j}^m \Gamma_{u,j}^m + \sigma^2\big)}_{\tilde{R}^1_{u,m}} \nonumber\\
&-\underbrace{\text{log}_{\text{2}}\big(\sum\limits_{\substack{i\in\mathcal{U} \\ i\neq u}}\sum\limits_{\substack{j\in\mathcal{D}}} p_{i,j}^m \Gamma_{u,j}^m + \sigma^2\big)}_{\tilde{R}^2_{u,m}}\Big) \geq \Omega, \quad \forall~ u  \label{R1R2}
\end{align}
\normalsize

From Equation (\ref{R1R2}), it can be noticed that this is a difference of two concave functions, i.e., $\tilde{R}^1_{u,m}$ and $\tilde{R}^2_{u,m}$, with respect to the UAV downlink power. The difference between two concave functions is not guaranteed to be neither concave nor convex. This motivates us to approximate $\tilde{R}^2_{u,m}$. To convert constraint (\ref{maxmin1}) to a convex one, we apply the SCA technique to approximate $\tilde{R}^2_{u,m}$ by a linear/convex function in each iteration. Let $p_{u,d}^m(r)$ is the given UAV downlink power in the r-th iteration. Since any concave function is globally upper-bounded by its first-order Taylor expansion at any point \cite{RuiZhangTraj}. Thus, the second term of Eq. (\ref{R1R2}), i.e., $\tilde{R}^2_{u,m}$, can be upper bounded as follows:

\small
\begin{align}
\tilde{R}^2_{u,m}=&\text{log}_{\text{2}}\big(\sum\limits_{\substack{i\in\mathcal{U} \\ i\neq u}}\sum\limits_{\substack{j\in\mathcal{D}}} p_{i,j}^m \Gamma_{u,j}^m + \sigma^2\big)\nonumber\\
\leq & \sum\limits_{\substack{i\in\mathcal{U} \\ i\neq u}}\sum\limits_{\substack{j\in\mathcal{D}}} \frac{\text{log}_{\text{e}}\Gamma_{u,j}^m}{\sum\limits_{\substack{i\in\mathcal{U} \\ i\neq u}}\sum\limits_{\substack{j\in\mathcal{D}}} p_{i,j}^m(r) \Gamma_{u,j}^m + \sigma^2} (p_{u,d}^m-p_{u,d}^m(r))\nonumber\\
+&\text{log}_{\text{2}}\big(\sum\limits_{\substack{i\in\mathcal{U} \\ i\neq u}}\sum\limits_{\substack{j\in\mathcal{D}}} p_{i,j}^m(r) \Gamma_{u,j}^m + \sigma^2\big)\nonumber\\
\overset{\Delta}{=}& \tilde{\tilde{R}}^2_{u,m}
\end{align}
\normalsize

Hence, constraint (\ref{maxmin1}) is now convex and it can be written as follows:

\small
\begin{equation}\label{8bconvex}
  \sum_{m} \Big(\text{log}_{\text{2}}\big(\sum\limits_{\substack{i\in\mathcal{U}}}\sum\limits_{\substack{j\in\mathcal{D}}} p_{i,j}^m \Gamma_{u,j}^m + \sigma^2\big)-\tilde{\tilde{R}}^2_{u,m}\Big) \geq \Omega(r)
\end{equation}
\normalsize

where $\Omega(r)$ is $\Omega$ at the r-th iteration. After converting constraint (\ref{maxmin1}) to a convex constraint, $\bf{P3}$ is now a convex optimization problem which can be solved efficiently.

\subsection{UAV Placement}

%Ahmed Alsahroa

In this subsection, we consider optimizing the UAVs' locations for fixed UAV association, resource and power allocations. Due to the non-convexity of the problem even with fixed association, resource and power allocations, we introduce an efficient algorithm to find the optimal UAVs' placement $\textbf{J}_d$.

The algorithm starts by dividing the desired area into equal sectors based on the number of the UAVs and each UAV is placed initially in the middle of the sector.
Initially, we generate certain number of particles in each sector to identify promising candidates and to form initial populations. Then, it determines the objective function achieved by selected particles by solving \textbf{P3}. After that, it finds the particle that provides the highest solution for this iteration.
Then, we generate a subset number of particles around this highest solution and calculate the objective function to find the best particle. This procedure is repeated until convergence or reach maximum iteration. To simplify the idea, this algorithm finds a candidate point among a large grid covering the disaster area. Hence, it finetunes by searching among a smaller grid surrounding each candidate point of the large grid until it finally finds the sub-optimal point which is the best point to minimize the objective function of \textbf{P3}.
%Fig.~\ref{Algorithm} shows an example iof 3 UAVs where the drone first will choose the best black dot location which maximize the objective function of \textbf{P2} then  it will finetune its location by searching for the best location among the gray points associated with the selected black dot.

%\begin{figure}[t!]
%  \centerline{\includegraphics[width=2.3in, height=1.7in]{Algorithm.pdf}}
%   \caption{\,: UAV placement}\label{Algorithm}
%\end{figure}

The following algorithm is used to solve \textbf{P2} by jointly solving \textbf{P3} for fixed coordinates and then finding the sub-optimal placement of the cDBSs.

%%%%%%%%%%%%%%%%%%%%%%%%%%%%%%%%%%%%%%%%%%%%%%%%%%%%
%%%%%%%%%%%%%%%%%% Algorithm 1 %%%%%%%%%%%%%%%%%%%%%
%%%%%%%%%%%%%%%%%%%%%%%%%%%%%%%%%%%%%%%%%%%%%%%%%%%%

\begin{algorithm}[h!]
\small
\caption{Joint optimization algorithm}\label{joint}
\KwIn{ Initial positions for UAVs $\mathbf{J}_d(0)$}
\KwOut{$\mathbf{J}_d(r+1),~ p_{u,d}^m(r+1),~ \Phi_{u,d}^m(r+1)$}
\begin{algorithmic}[1]
\WHILE {{Not} converged or reach maximum iteration}
\STATE Solve \bf{P3} for the given $\mathbf{J}_d(r)$
\STATE Denote results as $p_{u,d}^m(r+1)$ and $\Phi_{u,d}^m(r+1)$
\STATE Generate initial population ${\mathcal L}$ composed of $L$ particles
\FOR {$l=1 \cdots L$}
\STATE Compute corresponding objective function of \bf{P2} \\given $p_{u,d}^m(r+1)$ and $\Phi_{u,d}^m(r+1)$
%\STATE Find UAVs' bandwidth and power allocations by solving the optimization problem given in .... for all possible combinations ($d^L$).
\ENDFOR
\STATE Find $(l^{\text{r,local}}_d)=\underset{l,d}{\arg\mathrm{max}}~ \Omega^l - \sum_{d}\sum_{u}\sum_{m} p_{u,d}^m(l)$
%(i.e., $l^{\text{i,local}}_d$ indicates the index of the best local particle that results in the highest objective function for iteration $r$).
\STATE Generate a subset of particles around $l^{\text{r,local}}_d$
\STATE Use shrink-and-realign sample spaces process to find \\the best solution i.e., $l^{\text{r,sub-optimal}}_d$
\STATE $l^{\text{r,local}}_d=l^{\text{r,sub-optimal}}_d$, $\forall d$ \text{and} $\mathbf{J}_d(r+1)=l^{\text{r,sub-optimal}}_d$
\STATE Update r=r+1.
\ENDWHILE
\end{algorithmic}
\normalsize
\end{algorithm}

Algorithm 1 is an iterative efficient algorithm used to solve Problem $\bf{P2}$. Line 1 initiate the iteration and termination conditions then lines 2-3 solve $\bf{P3}$ for fixed UAVs' location. By fixing the placement of the UAVs and solving $\bf{P3}$ using successive convex approximation, then lines 4-7 generate particles and compute the objective function at each candidate point. From line 9 to 11 the algorithm finetunes the best placement by searching nearby for the best candidate and this is repeated at each iteration to find $l^{\text{r,local}}_d$ which indicates the index of the best local particle that results in the highest objective function for iteration $r$.

\section{Numerical Results}

In this section, numerical results are provided to investigate the benefits of using UAVs in mitigating GBSs failure in 5G networks. The simulation model consists of 1 failed GBS. We consider short-term and long-term failures in our simulation given one failure at a time; The multiple failures at the same time scenario is considered as a disaster which is a different problem. Under the short-term failure scenario, we initialized 4 standby DBSs to be used in the mitigation process. However, in the long-term failure scenario, we use only one helikite. Simulation was carried out using General Algebraic Modeling System (GAMS) \cite{gamss}. GAMS is a high-level modeling system for mathematical programming and optimization. It is designed for modeling and solving linear, nonlinear, and mixed-integer optimization problems. It consists of a language compiler and integrated high-performance solvers. GAMS is tailored for complex, large scale modeling applications, and allows to build large maintainable models that can be adapted quickly to new situations.

The simulation area is 400x400 $\text{m}^\text{2}$ where the failed BS is centered at the origin and the UEs of the failed BS are distributed randomly over this area. The UEs of the failed BS are static, however, the optimization problem is solved every time the distribution of the UEs is changed. The parameters used in the simulation are presented in Table II. Note that $h^s$ denotes the height of the short-term UAVs, i.e., drones, and $h^l$ denotes the height of the long-term UAVs, i.e., helikites.

\begin{table}
\centering
\vspace*{0.1in}
\caption{\label{tab2} System parameters}
%\hspace{-.7cm}
%\hspace{-.5cm}
\addtolength{\tabcolsep}{0.5pt}\begin{tabular}{|l|c||l|c|}
\hline
\textbf{Parameter}      & \textbf{Value}       & \textbf{Parameter}            & \textbf{Value}\\ \hline \hline
$P^{\text{max}}$ (W)    & $1$                  & $x_d^{\text{min}}$ (m)        & 0 \\[0.25ex]\hline
$R^{th}$ (bps/Hz)       & $0.5$                & $x_d^{\text{max}}$ (m)        & 400 \\[0.25ex] \hline
$h^{s}/h^{l}$ (m)       & $50/80$               & $y_d^{\text{min}}$ (m)        & 0\\[0.25ex] \hline
$p_{u,d}^m(r)$ (W)      & $0.1$                & $y_d^{\text{max}}$ (m)        & 400 \\[0.25ex] \hline
\end{tabular}
\end{table}

%\begin{figure}[!htb]
%            \centering
%        \includegraphics[width=3.5in, height=2.2in]{ShortLongRate.eps}
%        \caption{\,: Achievable rate versus maximum power}
%        \label{rates}
%\vspace{-0.1cm}
%\end{figure}

\begin{figure*}[!htb]
    \centering
    \begin{minipage}{3in}
        %\centering
        \centerline{\includegraphics[width=2.9in, height=2.1in]{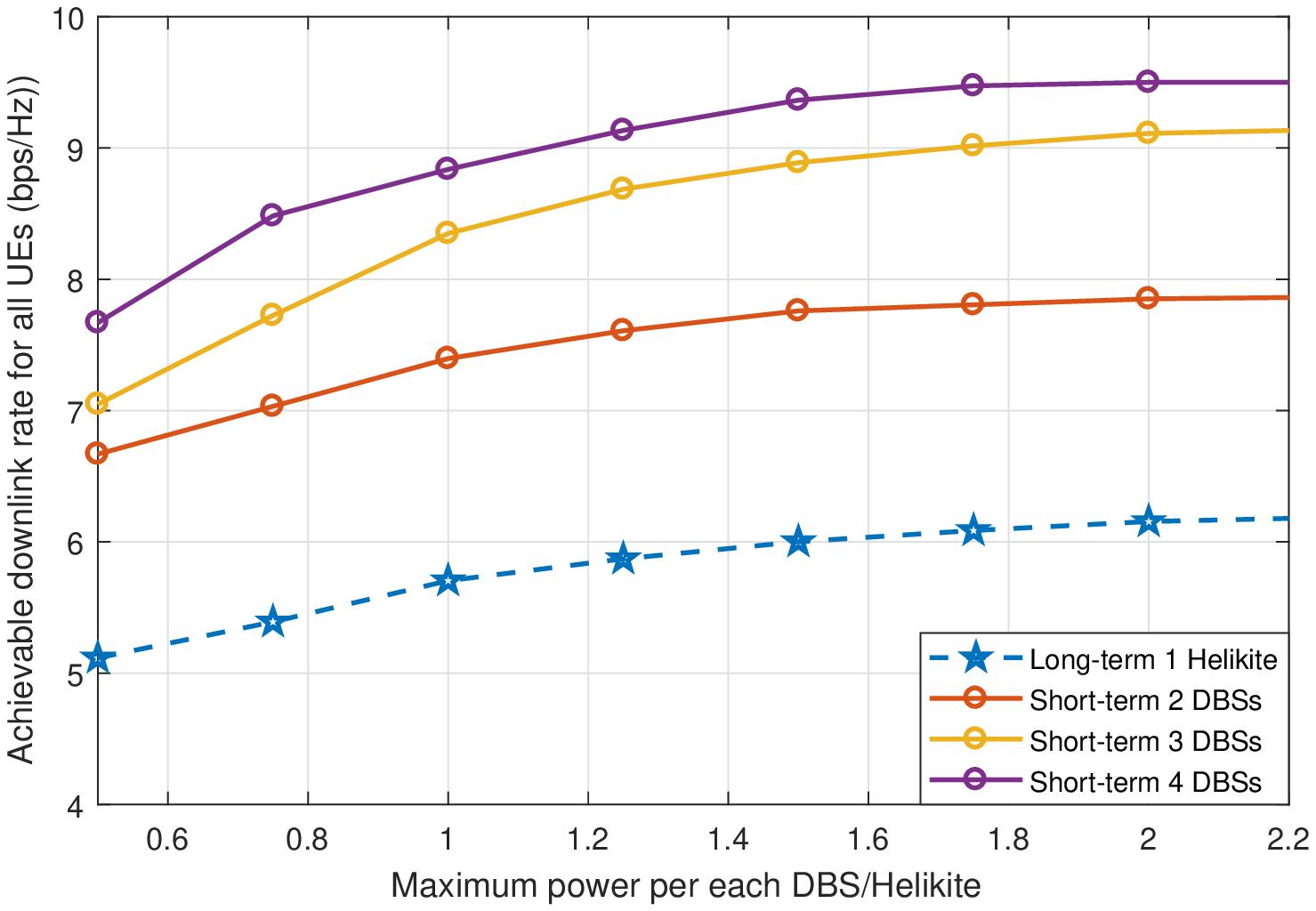}}
        \caption{\,: System model during normal operation.}
        \label{rates}
    \end{minipage}%
    \hspace*{0.3in}
    \begin{minipage}{3in}
        %\centering
        \centerline{\includegraphics[width=2.9in, height=2.1in]{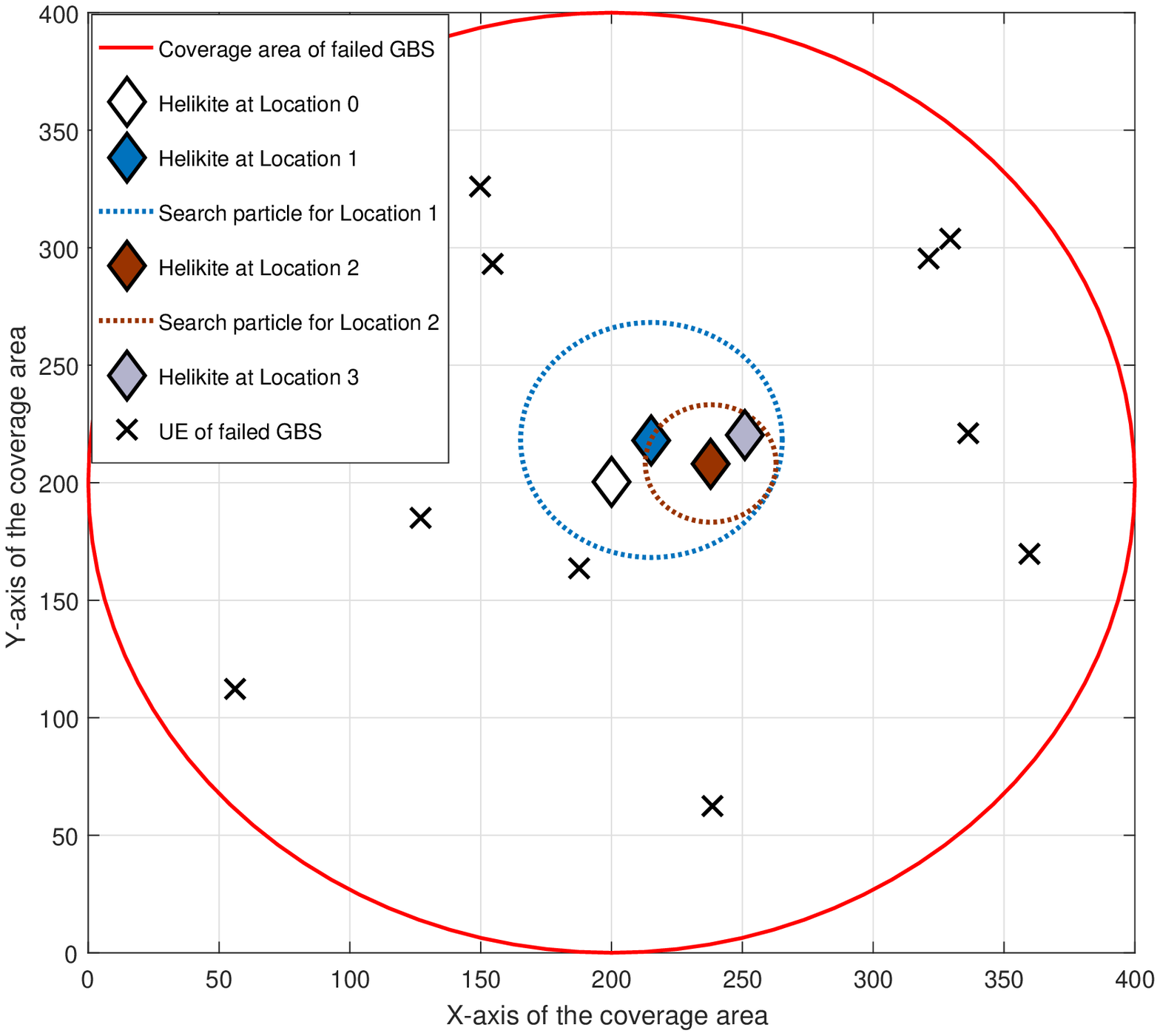}}
        \caption{\,: System model during one failure.}
        \label{locationss}
    \end{minipage}
\end{figure*}

In Fig. \ref{rates}, we present the short-term and long-term failure mitigation performance by plotting the achievable downlink rate of all UEs versus the maximum power per DBS/Helikite, i.e., $P^{\text{max}}$, in addition to varying the number of used DBSs in the short-term scenario for the same UEs' distribution and the threshold rate, i.e., $R^\text{th}$. However intuitively increasing the number of used DBSs consumes more power for hovering and hardware, As $P^{\text{max}}$ increases, the achievable rate of the UEs increases but levels of as the power reaches 1 W. This is because the objective function consists of two parts: 1) maximizing the minimum rate which guarantees fairness among all UEs, and 2) minimizing the downlink power of the DBSs/Helikite. It is worth noting that the excess power is only used to achieve the minimum rate requirement $R^\text{th}$.

The long-term scenario using 1 helikite results in the smallest rate. This is because the helikite altitude $h^{\text{l}}$ is greater than the DBSs altitude $h^{\text{s}}$ which consequently suffers from signal attenuation. Also, in the simulation and for comparison purposes, the maximum power of the helikite is set equal to that of the DBS. In reality, the helikite uses higher power levels, hence achieving higher rates.

Table III shows the association and UE power for both short-term and long-term scenarios. In case of long-term failure, the maximum power, $P^\text{max}$, assigned to the helikite is 2.25 W. Since in this scenario we are using only one helikite, it is obvious that there is a high variety in power levels among different UEs. For example, UE3 has the least power, 0.108 W, and this implies that this UE is near to the helikite. Furthermore, UE8 and UE2 use around 40\% of the helikite maximum power and this only happens in the case of the long-term failure since the helikite is covering the whole area of the failed GBS, hence satisfying the minimum rate of the far located UEs by increasing their transmission power. It is worth noting that the helikite is using its maximum power to serve its users. This implies that although minimizing the power as one of the objectives of the optimization problem, the helikite still must satisfy the minimum rate requirements of all UEs where the main objective of the optimization problem is to find the best location that helps in satisfying the rate constraint of the UEs by using the minimum power.

In case of short-term failure, although there are 4 DBSs available/standby, only GBS1, GBS2 and GBS4 are used to serve all UEs as shown in Table III. Given that the maximum power for each DBS is 1 W, DBS1 and DBS2 utilize less than 50\% of their maximum power since in this scenario not all UEs are associated with one UAV compared to the long-term scenario. The remaining power is not used since the minimum rate is already achieved beside the power minimization term used in the objective of the optimization problem. On the other hand, DBS4 utilized around 95\% of its maximum power. The reason for that is that half of the UEs are associated with this DBS. If the number of UEs increased or if the threshold rate is raised, then the last DBS, i.e., DBS3, will start to be involved and then the optimization problem will be solved again.

\small
\begin{table}
  \centering
  \vspace*{0.15in}
  \caption{\label{tab3} Association and power for 10 UEs}
  \renewcommand{\arraystretch}{1.2}
  \begin{tabular}{|p{1cm}|c|c|c|c|}
    \hline
    \multirow{2}{1cm}{\textbf{UEs}} & \multicolumn{2}{c|}{\textbf{Short-term}} & \multicolumn{2}{c|}{\textbf{Long-term}}\\
    % \hline
    % \textbf{Inactive Modes} & \textbf{Description}\\
    \cline{2-5}
    & \textbf{Association} & {$p_{u,d}$ (W)} & \textbf{Association} & {$p_{u,d}$ (W)} \\
    %\hhline{~--}
    \hline
    UE1 & GBS4 & 0.156 & Helikite  &0.176\\ \hline
    UE2 & GBS1 & 0.147 & Helikite  &0.397 \\ \hline
    UE3 & DBS4 & 0.105 & Helikite  &0.108 \\ \hline
    UE4 & GBS2 & 0.197 & Helikite  &0.203\\ \hline
    UE5 & DBS4 & 0.130 & Helikite  &0.239\\ \hline
    UE6 & GBS1 & 0.132 & Helikite  &0.115 \\ \hline
    UE7 & GBS4 & 0.171 & Helikite  &0.279 \\ \hline
    UE8 & GBS2 & 0.121 & Helikite  &0.451\\ \hline
    UE9 & DBS4 & 0.164 & Helikite  &0.153 \\ \hline
    UE10& GBS1 & 0.139 & Helikite  &0.129\\ \hline
  \end{tabular}
\vspace{-0.6cm}
\end{table}
\normalsize

Finally, Fig. \ref{locationss} investigates the long-term failure and its mitigation using one helikite. As shown in Algorithm 1, the initial position of the helikite is chosen to be in the center which is called location 0. Then the algorithm will find the best candidate location from $l$ locations which is named as location 2. A new search area of radius 50m centered at location 2 is used to find the best candidate location and the same approach repeated to find the finetuned location, i.e., location 3, which is considered to be the near optimal placement of the helikite.

\section{Conclusion}
In this paper, we proposed a novel self-healing framework for 5G networks assisted by two different types of UAVs to mitigate or at least alleviate the effect of any Ground base station (GBS) failure either if it is long-term or short-term failure. An optimization problem is formulated where its objective is to maximize the minimum achievable rate of the UEs under the failed BS by finding the optimal 2D placement of the UAVs in addition to minimizing the UAVs' downlink power.
%The UEs are served immediately after failure detection by the conventional self-healing technique until  UAVs arrive to their initial positions.

Results show that the minimum rate requirement is guaranteed for each UE under the failed BS. In addition, fairness is guaranteed among them where the minimum achievable rate is maximized for all UEs. The behavior of UAVs shows that each UAV is detecting its 2D location to serve its UEs based on the minimum rate requirement, i.e., $R^\text{th}$. These results show the ability of self-healing framework to mitigate either long-term or short-term failures of any GBS in the upcoming 5G networks. Addressing multi-GBS failures and using realistic channel model which considers the probability of line-of-site are an interesting future research direction.

%\begin{figure}[!htb]
%            \centering
%        \includegraphics[width=2.6in, height=2in]{locationsfinal.eps}
%        \caption{\,: DBS's and UEs' locations during 4 iterations.}
%        \label{locationss}
%\end{figure}

%in results
%From LOS Prob, we can see that by increasing the elevation angle or increasing the UAV altitude, the LOS prob increases.

\end{document}